\documentclass[prd,amsmath,twocolumn,10pt,superscriptaddress,floatfix,nofootinbib]{revtex4-2}

\usepackage[utf8]{inputenc}
\usepackage{amsmath,amssymb,amsfonts,bm}
\usepackage{epstopdf}
\usepackage{graphicx}
\usepackage{xcolor}
\usepackage{bbm}
\usepackage{multirow}
\usepackage{easyReview}
\usepackage{mathtools}
\usepackage{dsfont}
\usepackage{inputenc}
\usepackage{gensymb}
\usepackage{appendix}
\usepackage[
colorlinks=true,
linkcolor=magenta,
breaklinks=true,
urlcolor=blue,
citecolor=blue]{hyperref}

\graphicspath{{figures/}}

\newcommand{\itp}{\affiliation{Institute of Theoretical Physics, Chinese Academy of Sciences, Beijing 100190, China}}

\newcommand{\ucas}{\affiliation{School of Physical Sciences, University of Chinese Academy of Sciences, Beijing 100049, China}}

\newcommand{\scnt}{\affiliation{Southern Center for Nuclear-Science Theory (SCNT), Institute of Modern Physics, Chinese Academy of Sciences,\\ Huizhou 516000, China}}

\newcommand{\sxu}{\affiliation{College of Physics and Electronic Engineering, Shanxi University, Taiyuan 030006, China}}

\newcommand{\disc}{{\mathds{D}}}
\newcommand{\dgen}{{\mathds{H}}}
\newcommand{\conk}{{\mathds{K}}}
\newcommand{\cgen}{{\mathds{T}}}
\newcommand{\ppm}{{\mathds{P}}}
\newcommand{\tm}{{\bm{T}}}
\newcommand{\gm}{{\bm{G}}}
\newcommand{\hhm}{{\bm{h}}}
\newcommand{\iim}{{\bm{I}}}

\begin{document}
	
	\title{Discontinuity calculus and applications to two-body coupled-channel scattering}
	\author{Hao-Jie Jing}
	\email{jinghaojie@sxu.edu.cn}
	\sxu
	\author{Xiong-Hui Cao}
	\email{xhcao@itp.ac.cn}
	\itp 
	\author{Feng-Kun Guo}
	\email{fkguo@itp.ac.cn}
	\itp \ucas \scnt

	\date{\today}

	\begin{abstract}
		
		We present a novel method, termed discontinuity calculus, for computing discontinuities of complex functions. This framework enables a systematic investigation of both analytic continuation and the topological structure of Riemann surfaces. We apply this calculus to analyze the analytic continuation of partial-wave amplitudes in two-body coupled-channel scattering problems and discuss their uniformization of the corresponding Riemann surfaces. This methodology offers new perspectives and tools for analyzing coupled-channel scattering problems in quantum scattering theory. 
		
	\end{abstract}
	
	\maketitle
	
	\section{Introduction}
	\label{sec:intro}

	The analytic $S$-matrix theory is one of the most powerful tools for studying scattering problems. $S$-matrix elements, or scattering amplitudes, are functions of the invariant masses involved in reaction processes. These elements provide the invariant mass spectral distributions that can be directly compared with experimental data. In quantum field theory, bound or resonance states correspond to  poles in scattering amplitudes. 
	In hadron physics, particularly, the locations and residues of the poles contain crucial information about hadron properties; e.g., the branching fractions of decays of a resonance can be unambiguously defined through the residues of the poles~\cite{Heuser:2024biq}. 
	Causality constrains the unstable resonance poles to be on unphysical Riemann sheets of the complex energy plane, preventing their appearance on the physical sheet. Thus, it is necessary to perform analytic continuation of $S$-matrix elements, extending their domains from the real axis to the complex plane, to locate such poles.
	
	In this work, we propose a novel method, termed discontinuity calculus, for analytic continuation of complex functions and visualization of their Riemann surface structures. This approach provides a systematic way to derive the analytic continuation of any complex function along its branch cuts, as well as the structure of the corresponding Riemann surface. By applying this methodology, we have analyzed the analytic continuation and Riemann surface structure of the partial-wave scattering matrix in two-body coupled-channel problems~\cite{Oller:2019opk,Doring:2025sgb}. Furthermore, we establish the connection between the uniformization problem of the partial-wave scattering matrix and the uniformization theorem in complex analysis.
	
	This paper is organized as follows. In Sec.~\ref{sec:def-disc}, we introduce the discontinuity calculus. In Sec.~\ref{sec:2PGreenfunction}, we employ this framework to systematically investigate the analytic continuation of the scalar two-point Green's function and the topological structure of the associated Riemann surface. In Sec.~\ref{sec:T-matrix-RS}, we extend this analysis to the analytic continuation and Riemann surface topology of the partial-wave scattering matrix in two-body coupled-channel problems. In Sec.~\ref{sec:uniformization}, we address the uniformization problem of the partial-wave scattering matrix. Finally, Sec.~\ref{sec:summary} provides a summary.

	\section{Discontinuity calculus}
	\label{sec:def-disc}
	
	In this section, we develop a formalism for computing discontinuities of complex functions. 
	
	\subsection{Definitions}
	
	For any complex function $f(z)$, $F(\omega)$, and constant $\alpha$, we define the \emph{discontinuity operator} $\disc$, which complies with the following properties:\footnote{In Appendix~\ref{appx:def-dgen}, we provide an equivalent definition in terms of the operator $\dgen\equiv1-\disc$, for which the Leibniz and chain rules take simpler forms.}\\
	(1) Vanishing discontinuity of a holomorphic function $h(z)$:
	\begin{equation}
		\disc h(z)=0.\tag{R1}
		\label{eq:disc-rule-1}
	\end{equation}
	(2) Linearity:
	\begin{equation}
		\disc\left[\alpha_1 f_1(z)+\alpha_2 f_2(z)\right]=\alpha_1\disc f_1(z)+\alpha_2\disc f_2(z).\tag{R2}
		\label{eq:disc-rule-2}
	\end{equation}
	(3) Leibniz rule for the discontinuity of a product of two functions:  
	\begin{align}
		\disc \left[f_1(z)f_2(z)\right]=~&\disc f_1(z)~ f_2(z)+f_1(z)~\disc f_2(z)-\notag\\
		&\disc f_1(z)~\disc f_2(z).\tag{R3}
		\label{eq:disc-rule-3}
	\end{align}
	(4) Chain rule for the discontinuity of a composite function $F\left[f(z)\right]$:
	\begin{align}
		\disc F\left[f(z)\right]=~&F\left[f(z)\right]-\big[F\left(\omega\right) - \disc F(\omega)\big]_{\omega=f(z)-\disc f(z)}.\tag{R4}
		\label{eq:disc-rule-4}
	\end{align}
	In particular, if $\disc F(\omega) = \disc f(z) = 0$, then $\disc F\left[f(z)\right] = 0$; if $\disc F(\omega) = 0$ and $\disc f(z) \neq 0$, then $\disc F\left[f(z)\right]=F\left[f(z)\right]-F\left[f(z)-\disc f(z)\right]$; if $\disc F(\omega) \neq 0$ and $\disc f(z) = 0$, then $\disc F\left[f(z)\right]=\disc F(\omega)\big|_{\omega=f(z)}$.

	Using (\ref{eq:disc-rule-1},\ref{eq:disc-rule-3}), one can prove that the discontinuities of the functions $f(z)$ and $1/f(z)$ satisfy the following relation:
	\begin{equation}
		\disc f(z)=-f(z)\,\disc [1/f(z)]\left[f(z)-\disc f(z)\right].
		\label{eq:disc-f-and-f-1}
	\end{equation}
	
	In general, the discontinuity of a complex function $f(z)$ can be decomposed into the following form:
	\begin{equation}
		\disc f(z) =\sum_{i=1}^{n}~ \conk_i f(z) ~\theta_i(z)+\sum_{j=1}^{n'}~\alpha_j\delta_j(z).
		\label{eq:def-conk}
	\end{equation}
	The right-hand side of Eq.~\eqref{eq:def-conk} comprises two types of contributions: branch cuts and poles.
	The first term describes the branch cuts of $f(z)$, which are generally expressed using the Heaviside $\theta$ function. Here, they are represented by $\theta_i(z)~(i=1,\cdots,n)$ in simplified notation, where $n$ denotes the number of cuts. The specific form of the argument within the $\theta$ function depends on the particular cut. The components $\conk_i f(z)$ determine the analytic continuation of $f(z)$. 
	In the second term, the components $\delta_j(z)~(j=1,\cdots,n')$ correspond to the poles of $f(z)$, where $n'$ is the number of poles. The discontinuity arising from poles is typically expressed using the Dirac $\delta$ function and its derivatives; the coefficients $\alpha_j$ are determined by the Laurent series expansion of $f(z)$.
	We refer to $\conk_i f(z)$ as the \emph{continuation kernel} of $f(z)$, and $\conk_i$ as the continuation kernel operator. 
	For a given cut of a complex function $f(z)$ or a composite function $F\left[f(z)\right]$, represented by $\theta_i(z)$, it can be readily shown that the corresponding continuation kernel operator $\conk_i$ also satisfies (\ref{eq:disc-rule-1}--\ref{eq:disc-rule-4}).

	Furthermore, we can define the \emph{continuation generator} $\cgen_i \equiv 1 - \conk_i$. The action of the continuation generator $\cgen_i$ on the function $f(z)$, denoted $\cgen_i f(z)$, yields the expression of $f(z)$ after analytic continuation across the cut specified by $\theta_i(z)$. 
	We can then calculate the discontinuity of the function $\cgen_i f(z)$:
	\begin{equation}
		\disc\left[\cgen_i f(z)\right]=\sum_{j=1}^{n}\conk_{j}\left[\cgen_i f(z)\right]\theta_{j}(z)+\sum_{j=1}^{n'}~\alpha_j\delta_j(z),
		\label{eq:disc-cgen}
	\end{equation}
	which leads to the expression of the function $\cgen_i f(z)$ after analytic continuation across the cut of $\cgen_i f(z)$ specified by $\theta_j(z)$, namely $\cgen_{j}\cgen_{i} f(z)$.
	By repeating this process, we can obtain the expression of $f(z)$ on all Riemann sheets after analytic continuation, and thus determine the structure of the Riemann surface $\mathrm{RS}\{f(z)\}$. 
	
	\subsection{Examples}

	Let us now examine some instructive examples.
	
	{\bf Example 1}: the square root function $f(z)=\sqrt{z}$ $(z \in \mathbb{C})$.
	Using \eqref{eq:disc-rule-3} to expand $\disc f^2(z)=0$, the following equation is obtained:
	\begin{align}
		\left(\disc \sqrt{z}-2 \sqrt{z}\right) \disc \sqrt{z}=0.
		\label{eq:disc-sqrt-0}
	\end{align}
	The square root function has two branch points at 0 and $\infty$, and any continuous curve connecting these branch points can serve as a branch cut. When the branch cut is chosen along the positive real axis, the following equation holds:
	\begin{equation}
		\disc\sqrt{z}=2\sqrt{z}\times\theta(z)+0\times\theta(-z)=2\sqrt{z}~\theta(z),
		\label{eq:disc-sqrt}
	\end{equation}
	where the two terms following the first equality correspond to the two solutions of Eq.~\eqref{eq:disc-sqrt-0}.
	Thus, the continuation kernel for the square root function is $\conk_1 \sqrt{z}=2 \sqrt{z}$, with the trivial case $\conk_2\sqrt{z}=0$ omitted. 
	
	The continuation generator is given by $$\cgen_1 \sqrt{z}=\left(1-\conk_1\right) \sqrt{z}=-\sqrt{z}.$$ 
	Since $\cgen_1 \sqrt{z}$ and $\sqrt{z}$ differ only by a sign, it follows that $\disc\left(\cgen_1 \sqrt{z}\right)=-\disc \sqrt{z}$. Therefore, the analytic continuation of $\cgen_1 \sqrt{z}$ does not introduce new continuation generators. Furthermore, since $\cgen_1^2 \sqrt{z}=-\cgen_1 \sqrt{z}=\sqrt{z}$, this indicates that the order of $\cgen_1$ is 2, generating a cyclic group of order 2, denoted $\mathbb{Z}_2$. 
	
	In summary, the Riemann surface of the square root function $\operatorname{RS}\{\sqrt{z}\}$ consists of two Riemann sheets:
	$$
	\operatorname{RS}\{\sqrt{z}\}=\mathbb{C} \times\left\{\cgen_1 \mid \cgen_1^2 \sim 1\right\} \cong \mathbb{C} \times \mathbb{Z}_2.
	$$
	Moreover, it can also be easily shown that $\operatorname{RS}\{\sqrt[n]{z}\} \cong \mathbb{C} \times \mathbb{Z}_n$ for $n \in \mathbb{N}^*$.
	
	
	{\bf Example 2}: the logarithmic function $f(z)=\log z~(z \in \mathbb{C}^* \equiv \mathbb{C} \backslash\{0\})$.
	Using \eqref{eq:disc-rule-4} to expand $\disc e^{\log z}=0$, the following equation is obtained:
	$$
	e^{\disc \log z}=1.
	$$ 
	The logarithmic function also has two branch points located at 0 and $\infty$. If the branch cut is chosen along the negative real axis, the above equality leads to the discontinuity of the logarithm:
	$$
	\disc \log z=2 m \pi i \theta(-z)\quad (m \in \mathbb{Z}).
	$$
	From this expression, the continuation kernel is given by $\conk_m \log z=2 m \pi i$.
	
	The continuation generator is given by 
	$$\cgen_m \log z = (1-\conk_m) \log z=\log z-2 m \pi i.$$ 
	Since $\disc(\conk_m \log z)=0$, no new continuation generators are needed for the analytic continuation of $\cgen_m \log z$. Furthermore, group multiplication is expressed as
	$
	\cgen_{m_1}\left(\cgen_{m_2} \log z\right)=\cgen_{m_1+m_2} \log z.
	$
	This result establishes that the continuation generators $\cgen_{ \pm 1}$ alone suffice to construct the complete analytic continuation. Consequently, the discontinuity of the logarithmic function can be rewritten as:
	\begin{equation}
		\disc\log{z}=\pm2\pi i\,\theta(-z).
		\label{eq:disc-log}
	\end{equation}
	The sign $+(-)$ on the right-hand side corresponds to the result obtained by analytic continuation of the logarithmic function along the upper (lower) edge of the branch cut. 
	
	In summary, the Riemann surface of the logarithmic function, $\operatorname{RS}\{\log z\}$, consists of countably infinite Riemann sheets and can be expressed as:
	$$
	\operatorname{RS}\{\log z\}=\mathbb{C}^* \times\left\{\cgen_1, \cgen_{-1} \mid \cgen_1 \cgen_{-1} \sim 1\right\} \cong \mathbb{C}^* \times \mathbb{Z}.
	$$
	
	{\bf Example 3}: the meromorphic function $f(z)=1/z$ $(z \in \mathbb{C}^*)$. By taking the derivative of both sides of Eq.~\eqref{eq:disc-log} with respect to $z$ and interchanging the order of the derivative operator and the discontinuity operator, which is valid due to the linearity property~\eqref{eq:disc-rule-2}, one obtains:
	\begin{equation}
		\disc\left(\frac{1}{z}\right)=\mp\,2\pi i\,\delta(z),
		\label{eq:disc-1/x}
	\end{equation}
	where the sign $-(+)$ corresponds to the variable $z$ approaching zero along the upper (lower) edge of the real axis, namely $\operatorname{Im}z>0$ $(\operatorname{Im}z<0)$. 
	Eq.~\eqref{eq:disc-1/x} indicates that the function $1/z$ has only a single pole at $z=0$.
	Furthermore, we have the following identities $(n\in\mathbb{N})$:
	\begin{equation}
		\disc\left(\frac{1}{z^{n+1}}\right)=\mp\,2\pi i\,\frac{(-1)^{n}}{n!}\frac{\mathrm{d}^n}{\mathrm{d}z^n}\delta(z).\notag
	\end{equation} 
	
	\section{Riemann surface of two-point Green's function} 
	\label{sec:2PGreenfunction}
	
	In this section, we employ the discontinuity calculus formalism developed in Sec.~\ref{sec:def-disc} to systematically investigate the Riemann surface topology of the scalar two-point one-loop Green's function,
	\begin{equation}
		G\left(p^2\right)\equiv\int_{\mathbb{R}^4}\frac{i\mathrm{d}^4q}{(2\pi)^4}\frac{1}{\left(q_+^2-m_1^2+i\epsilon\right)\left(q_-^2-m_2^2+i\epsilon\right)},
		\label{eq:def-2PGreenfunction}
	\end{equation}
	where $q_\pm=p/2\pm q$, $p$ is the four-momentum of the external line, $m_1$ and $m_2$ are the masses of the intermediate particles, and $\epsilon=0^{+}$ is a positive infinitesimal, which is used to determine the way of integration path bypassing the poles of the integrand.
	
	First, we demonstrate that the discontinuity of the Green’s function $G(s)~(\text{where }s=p^2)$ can be explicitly derived solely from its definition in Eq.~\eqref{eq:def-2PGreenfunction}. By applying the discontinuity calculus in Eq.~\eqref{eq:disc-1/x}, the discontinuity of $G(s)$ in the physical region, which is defined as the upper edge of the real axis on the first Riemann sheet of the complex $s$ plane, can be obtained from the following substitution in Eq.~\eqref{eq:def-2PGreenfunction}:\footnote{Employing the discontinuity calculus, it is possible to determine not only the discontinuities of $G(s)$ in the physical region but also those in the unphysical region. A streamlined derivation of this procedure is presented in Appendix~\ref{appx:disc-2PGfunc-def}.} 
		\begin{align}
			&\frac{1}{\left(q_+^2-m_1^2+i\epsilon\right)\left(q_-^2-m_2^2+i\epsilon\right)}\mapsto\notag\\
			&\quad\quad \quad\quad\left[-2\pi i\delta_{\text{p}}\left(q_+^2-m_1^2\right)\right]\left[-2\pi i\delta_{\text{p}}\left(q_-^2-m_2^2\right)\right],\notag
		\end{align}
		where the subscript ``p" of the Dirac-$\delta$ functions means that only the contribution of the ``proper" root of $q_{+/- }^2=m_{1/2}^2$, for which the temporal component $q_{+/-}^0$ is positive, is to be taken. 
		The validity of this substitution rule, termed the Cutkosky rule~\cite{Cutkosky:1960sp}, can be immediately verified within the framework of discontinuity calculus. 
		It is plausible that, for any complex function $f(z)$, even without knowing its analytical expression, the complete analytic continuation and the topological structure of its Riemann surface can be obtained via discontinuity calculus, provided that the specific form of its discontinuity $\disc f(z)$ is known. 
		%
		This approach can be utilized to perform the analytic continuation of amplitudes with, e.g., triangle and box diagrams~\cite{Guo:2019twa,Shen:2025nen}.
		
		In what follows, we perform a systematic analysis of the discontinuities in the explicit expressions for $G(s)$ using dimensional regularization.
        In the dimensional regularization scheme, the scalar two-point one-loop integral $G(s)$ in Eq.~\eqref{eq:def-2PGreenfunction} can be expressed in the following form (see, e.g.,~\cite{Oller:1998zr,Yao:2020bxx}):
		\begin{align}
			G(s)=\frac{1}{(4\pi)^2}&\left[a(\mu)+\log\left(\frac{m_1m_2}{\mu^2}\right)+\right.\notag\\
			&\quad\quad\quad\quad\quad\left.\frac{\Delta}{s}\log{\frac{m_1}{m_2}}+8\pi L(s)\right],
			\label{eq:2PGfunc-DR-Int}
		\end{align}
		where $a(\mu)$ is a subtraction constant, and $\Delta=m_1^2-m_2^2$. The function $L(s)$ is given by
		\begin{equation}
			L(s)=\rho(s)\varphi(s),
			\label{eq:def-L}
		\end{equation}
		where\footnote{The value of the function $L(s)$ is independent of the double-valuedness of the square root in $\rho(s)$ since it remains invariant under the inversion transformation $\rho(s)\to-\rho(s)$. Nevertheless, to avoid ambiguity, we stipulate that the imaginary part of $\rho(s)$ is always positive.}
		\begin{align}
			\rho(s)=~&\frac{1}{16\pi s}\sqrt{(s-s_+)(s-s_-)},\notag\\
			\varphi(s)=~&\log\left[c-s-16\pi s\rho(s)\right]-\log\left[c-s+16\pi s\rho(s)\right],\notag
		\end{align}
		with $s_+=(m_1+m_2)^2$ and $s_-=(m_1-m_2)^2$ being the threshold and pseudo-threshold, respectively, and $c=(s_++s_-)/2=m_1^2+m_2^2$.
		
		Using (\ref{eq:disc-rule-1},\ref{eq:disc-rule-2}) and Eq.~\eqref{eq:disc-1/x}, the action of the discontinuity operator $\disc$ on the Green's function $G(s)$ is\footnote{The physical regime is defined by approaching the real axis from above ($\operatorname{Im}s \rightarrow 0^+$) along the interval $[s_+, \infty)$. Therefore, an analytic continuation along the upper edge of the real axis is considered. Specifically, when employing the result of Eq.~\eqref{eq:disc-log} (Eq.~\eqref{eq:disc-1/x}), the sign $+(-)$ is selected on the right-hand side.}
		\begin{equation}
			\disc G(s)=\frac{\Delta}{8\pi i}\log{\frac{m_1}{m_2}}\,\delta(s)+\frac{1}{2\pi}\disc L(s).
			\label{eq:disc-G1-mid}
		\end{equation}
		Then, according to \eqref{eq:disc-rule-3}, one has
		\begin{equation}
			\disc L(s)=\disc\rho(s)\,\varphi(s)+\rho(s)\disc\varphi(s)-\disc\rho(s)\disc\varphi(s).
			\label{eq:disc-L(s)-mid}
		\end{equation}
		Subsequently, we only need to calculate the discontinuities of $\rho(s)$ and $\varphi(s)$, respectively.
		For $\rho(s)$, based on (\ref{eq:disc-rule-1}, \ref{eq:disc-rule-3}, \ref{eq:disc-rule-4}) and Eqs.~(\ref{eq:disc-sqrt},\ref{eq:disc-1/x}), one can obtain
		\begin{equation}
			\disc\rho(s)=2\rho(s)\theta\left[(s-s_+)(s-s_-)\right]-\frac{\Delta}{8i}\,\delta(s).
			\label{eq:disc-rho}
		\end{equation}
		The other part $\varphi(s)$ is the difference between two logarithmic functions. For the first logarithmic function, according to (\ref{eq:disc-rule-1}--\ref{eq:disc-rule-4}) and Eqs.~(\ref{eq:disc-log},\ref{eq:disc-rho}), one can obtain:
		\begin{align}
			\disc\log{[c-s-16\pi s\rho(s)]}=~&\varphi(s)\theta\left[(s-s_+)(s-s_-)\right]+\notag\\
			&2\pi i\theta(s-s_+),
		\end{align}
		where the relation $\theta[s\pm16\pi s\rho(s)-c]=\theta\left(s-s_+\right)$ is utilized. 
		On the other hand, since 
		$$\log [c-s-16\pi s\rho(s)]+\log [c-s+16\pi s\rho(s)]\hspace{-.8mm}=\hspace{-.8mm}\log \left(c^2-\Delta^2\right)$$ is a constant, it follows that $$\disc\log [c-s+16\pi s\rho(s)]=-\disc\log [c-s-16\pi s\rho(s)].$$ 
		Consequently, one has
		\begin{equation}
			\disc\varphi(s)=2\varphi(s)\theta\left[(s-s_+)(s-s_-)\right]+4\pi i \theta(s-s_+).
			\label{eq:disc-varphi}
		\end{equation}
		Then, substituting Eqs.~(\ref{eq:disc-rho},\ref{eq:disc-varphi}) into Eq.~\eqref{eq:disc-L(s)-mid}, one obtains 
		$$\disc L(s)=-4 \pi i \rho(s) \theta\left(s-s_+\right)-\frac{\Delta}{4i}\log{\frac{m_1}{m_2}}\,\delta(s).$$
	Further substituting the above result into Eq.~\eqref{eq:disc-G1-mid} yields
	\begin{equation}
		\disc G(s)=-2i\rho(s)\theta\left(s-s_+\right).
		\label{eq:disc-G1}
	\end{equation}
	
\begin{figure}[tb]
		\centering
		\includegraphics[width=0.85\linewidth]{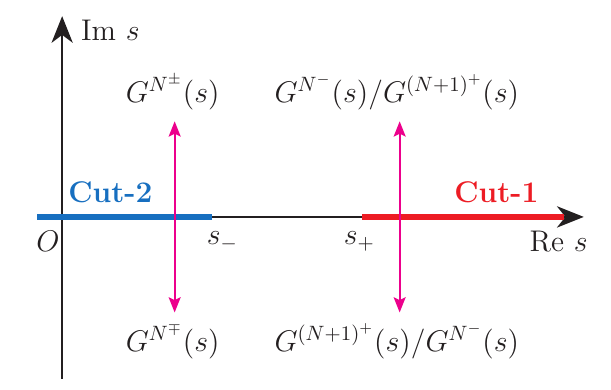}
		\caption{Schematic diagram illustrating the cuts of the Green's function $G(s)$ in the complex $s$-plane. The red line on the real axis represents Cut-1, while the blue line represents Cut-2. The two Riemann sheets corresponding to the Green's functions are interconnected along their respective cuts, as indicated by the pink arrows.}
		\label{fig:1}
	\end{figure}

	Equation~\eqref{eq:disc-G1} means that $G(s)$ has a branch cut, denoted as Cut-1, in the complex $s$-plane, starting from the threshold $s_+$ and extending to positive infinity; see the red line in Fig.~\ref{fig:1}. The continuation kernel of $G(s)$ along Cut-1 is $\conk_1 G(s)=-2 i \rho(s)$.
	
	The Riemann sheet corresponding to $G(s)$ is denoted as $\operatorname{RS}\{G(s)\}$-$\mathrm{I}$, which is called the physical sheet. All other Riemann sheets are called unphysical sheets. The unphysical sheet (lower half-plane) connected to the physical sheet (upper half-plane) along Cut-1 is denoted as $\operatorname{RS}\{G(s)\}$-$\mathrm{II}^{+}$. The corresponding Green's function $G^{\mathrm{II}^{+}}\!(s)$ can be obtained by applying the continuation generator $\cgen_1 = 1-\conk_1$ to $G(s)$:
	\begin{equation}
		G^{\mathrm{II}^+}\!(s)= \cgen_1G(s)= G(s)+2i\rho(s).
		\label{eq:def-G2p}
	\end{equation}

	Next, let us examine the discontinuity of $G^{\mathrm{II}^+}\!(s)$. Combining Eq.~\eqref{eq:disc-rho} with Eq.~\eqref{eq:disc-G1}, one obtains
	\begin{align}
			\disc\left[G^{\mathrm{II}^+}\!(s)\right]\hspace{-1.4mm}=\,&\disc G(s)+2i\disc\rho(s)\notag\\
			=\,&2i\rho(s)\theta(s-s_+)\hspace{-.8mm}+\hspace{-.8mm}4i\rho(s)\theta(s_--s)\hspace{-.8mm}-\hspace{-.8mm}\frac{\Delta}{4}\delta(s),
			\label{eq:disc-G2p}
	\end{align}
	where the relation $\theta[(s-s_+)(s-s_-)]=\theta(s-s_+)+\theta(s_--s)$ has been used.
		
	The above results indicate that there are two cuts (and a pole at $s=0$) on the Riemann sheet $\operatorname{RS}\{G(s)\}$-$\mathrm{II}^{+}$.
	One is Cut-1, with the corresponding continuation kernel $\conk_1 G^{\mathrm{II}^{+}}\!(s)=2 i \rho(s)$. The other, denoted as Cut-2, starts from the pseudo-threshold $s_-$ and extends toward negative infinity; see the blue line in Fig.~\ref{fig:1}. Notice that Cut-2 does not appear on the physical sheet. 
	Therefore, a new continuation kernel $\conk_2$ needs to be introduced, which satisfies $\conk_2 G(s)=0$ and $\conk_2 G^{\mathrm{II}^{+}}\!(s)=4 i \rho(s)$. The analytic continuations of $G^{\mathrm{II}^{+}}\!(s)$ along Cut-1 and Cut-2 are given by the actions of the continuation generators $\cgen_1$ and $\cgen_2=1-\conk_2$, respectively:
	\begin{align}
		\cgen_1\left[G^{\mathrm{II}^+}\!(s)\right]&=G^{\mathrm{II}^+}\!(s)-2i\rho(s)=G(s),\notag\\
		\cgen_2\left[G^{\mathrm{II}^+}\!(s)\right]&=G^{\mathrm{II}^+}\!(s)-4i\rho(s)= G^{\mathrm{II}^-}\!(s),
		\label{eq:cgen-G2}
	\end{align}
	where we have defined $G^{\mathrm{II}^-}\!(s)\equiv G(s)-2i\rho(s)$, and the corresponding Riemann sheet is denoted as $\mathrm{RS}\{G(s)\}$-$\mathrm{II}^-$. These results demonstrate that $\mathrm{RS}\{G(s)\}$-$\mathrm{II}^+$ (upper half-plane) is connected to the physical sheet (lower half-plane) along Cut-1; simultaneously, $\mathrm{RS}\{G(s)\}$-$\mathrm{II}^+$ (upper half-plane) is connected to $\mathrm{RS}\{G(s)\}$-$\mathrm{II}^-$ (lower half-plane) along Cut-2.
	
	Furthermore, the discontinuity of $G^{\mathrm{II}^{-}}\!(s)$ can be analyzed and its analytic continuation carried out.
	Since Eq.~\eqref{eq:cgen-G2} contains only $G(s)$ and $\rho(s)$, it follows from Eqs.~(\ref{eq:disc-rho},\ref{eq:disc-G1}) that further analytic continuation will not introduce new continuation kernels. The actions of the continuation generators $\cgen_1$ and $\cgen_2$ on $G(s)$ and $\rho(s)$ are summarized as follows:
	\begin{align}
		&\cgen_1 G(s)=G(s)+2i\rho(s),\quad\cgen_1 \rho(s)=-\rho(s),\notag\\
		&\cgen_2 G(s)=G(s),\quad\quad\quad\quad~~\,\cgen_2\rho(s)=-\rho(s).
		\label{eq:ex-3-cgen-summary}
	\end{align}
	From Eq.~\eqref{eq:ex-3-cgen-summary}, the following equations are obtained:
	\begin{align}
		(m\in\mathbb{N}):\quad&\left(\cgen_1\cgen_2\right)^mG(s)=G(s)+2mi\rho(s),\notag\\
		&\left(\cgen_2\cgen_1\right)^mG(s)=G(s)-2mi\rho(s).\notag
	\end{align}
	
	Based on the above discussions, it can be seen that the analytic continuation of $G(s)$ will generate countably infinite Riemann sheets $\operatorname{RS}\{G(s)\}$-$N^{ \pm}$. The Green's functions on the corresponding Riemann sheets can be defined as follows $(N=\mathrm{I}, \mathrm{II},\mathrm{III},\cdots)$:
	\begin{equation}
		G^{N^{\pm}}\!(s)\equiv G(s)\pm 2i\left(N-1\right)\rho(s).
		\label{eq:2PGfunc-cgen-res}
	\end{equation}
	In particular, $G^{\mathrm{I}^{ \pm}}\!(s)\equiv G(s)$. These Riemann sheets are connected to each other in the following ways:
	\begin{enumerate}
		\item[(1)] The upper (lower) half-plane of the Riemann sheet $\operatorname{RS}\{G(s)\}$-$N^{-}$ is connected to the lower (upper) half-plane of $\operatorname{RS}\{G(s)\}$-$(N+1)^{+}$ along Cut-1; 
		\item[(2)] The upper (lower) half-plane of the Riemann sheet $\operatorname{RS}\{G(s)\}$-$N^{+}$ is connected to the lower (upper) half-plane of the Riemann sheet $\operatorname{RS}\{G(s)\}$-$N^{-}$ along Cut-2. 
	\end{enumerate}

	These connection patterns can be summarized by the following algebraic expressions:
	\begin{align}
		&\cgen_1G^{N^-}\!(s)=G^{(N+1)^+}\!(s),\quad\cgen_1G^{(N+1)^+}\!(s)=G^{N^-}\!(s).\notag\\
		&\cgen_2G^{N^+}\!(s)=G^{N^-}\!(s),\quad\quad~~\cgen_2G^{N^-}\!(s)=G^{N^+}\!(s).\notag
	\end{align}
	An intuitive representation of the connection patterns among the Riemann sheets $\mathrm{RS}\{G(s)\}$-$N^{\pm}$ is shown in Fig.~\ref{fig:1}. In summary, the structure of the Riemann surface $\mathrm{RS}\{G(s)\}$ can be expressed as follows:
	\begin{equation}
		\mathrm{RS}\{G(s)\}=\mathbb{C}\times\left\{\cgen_1,\cgen_2\big|\cgen_1^2\sim1,\cgen_2^2\sim1\right\}\cong\mathbb{C}\times\mathbb{Z}.\notag
	\end{equation}

	\begin{figure*}[tbh] 
		\centering
		\includegraphics[width=0.34\linewidth]{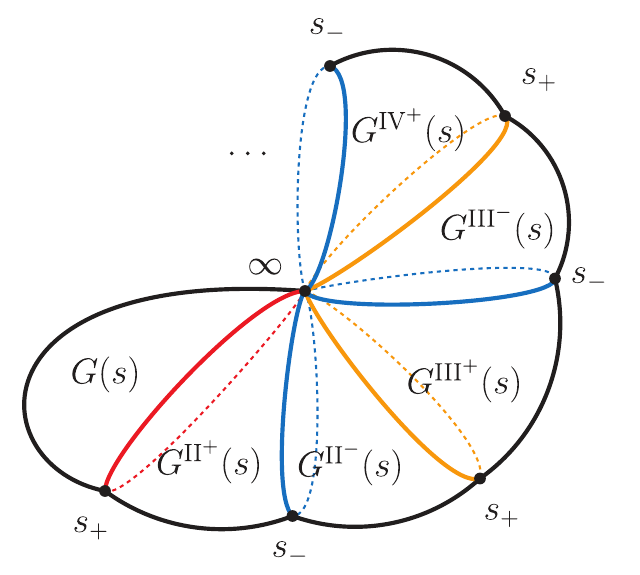}
		\includegraphics[width=0.35\linewidth]{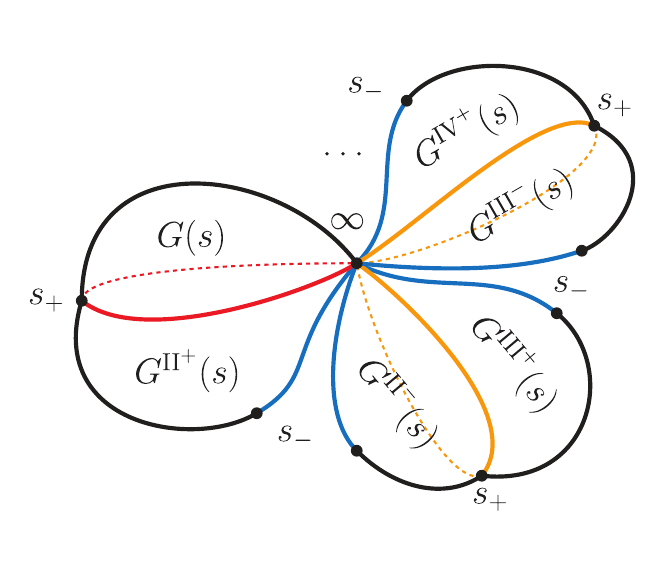}\quad\quad
		\includegraphics[width=0.21\linewidth]{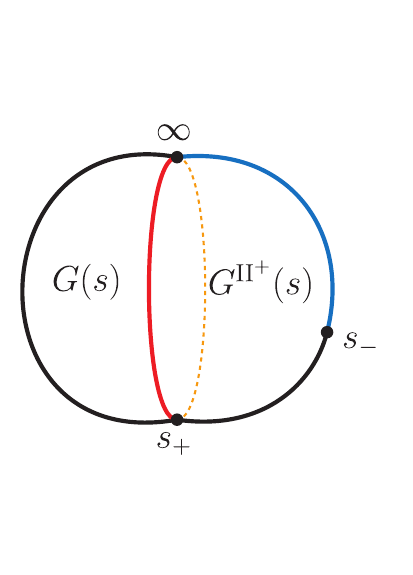}
		\caption{Left: Riemann surface $\overline{\operatorname{RS}}\{G(s)\}$, which consists of countably infinite Riemann sheets, with the corresponding Green's function labeled in the plot. Middle: Riemann surface $\overline{\operatorname{RS}}\{G(s)\}$ when the analytic continuation along Cut-2 is not considered; the originally simply connected Riemann surface is divided into multiple independent branches, each consisting of two Riemann sheets. Right: When the analytic continuation along Cut-2 is disregarded, the branch in the middle plot containing the physical sheet consists of two Riemann sheets; notably, the entire branch is topologically equivalent to a sphere.
		Here, the red lines represent Cut-1 on the physical sheet, the blue lines represent Cut-2 on the unphysical sheets, the orange lines represent Cut-1 on the unphysical sheets, and the black dots represent the branch points of the Green's functions. }
		\label{fig:2}
	\end{figure*}
	To more intuitively understand the topological structure of the Riemann surface $\operatorname{RS}\{G(s)\}$, let us extend the domain of $G(s)$ from the complex plane $\mathbb{C}$ to the Riemann sphere $\hat{\mathbb{C}}$. The Riemann surface obtained through analytic continuation on the Riemann sphere $\hat{\mathbb{C}}$ is denoted as $\overline{\operatorname{RS}}\{  G(s)\}$. The topological structure of $\overline{\operatorname{RS}}\{  G(s)\}$ can be visualized in the leftmost panel of Fig.~\ref{fig:2}.
	In Fig.~\ref{fig:2}, the red solid line represents Cut-1 on the physical sheet, i.e., the right-hand cut; the orange solid and dashed lines represent Cut-1 on unphysical sheets; the blue solid and dashed lines represent Cut-2 on unphysical sheets. The black dots indicate the branch points of the Green's functions. It can be clearly observed that the Riemann sheet $\operatorname{RS}\{G(s)\}$-$\mathrm{I}$ is connected to $\operatorname{RS}\{G(s)\}$-$\mathrm{II}^{+}$ via Cut-1 (the red line). Similarly, $\operatorname{RS}\{G(s)\}$-$\mathrm{II}^{+}$ is connected to $\operatorname{RS}\{G(s)\}$-$\mathrm{II}^{-}$ via Cut-2 (the blue line), while $\operatorname{RS}\{G(s)\}$-$\mathrm{II}^{-}$ is connected to $\operatorname{RS}\{G(s)\}$-$\mathrm{III}^{+}$ via Cut-1 (the orange line). Starting from the physical sheet, crossing Cut-1 and Cut-2 sequentially leads to successive transitions into new unphysical sheets, and this process can continue indefinitely.
	The Riemann surface $\overline{\operatorname{RS}}\{G(s)\}$ resembles a snail shell composed of individual ridges. Each ridge represents a Riemann sheet, the connections between adjacent ridges correspond to the cuts, and the central point where all connections converge corresponds to the point at infinity.

	Before closing this section, we analyze a key special case: the Riemann surface structure of the Green's function $G(s)$ when analytic continuation across Cut-2 is neglected.
	From an algebraic perspective, this scenario is equivalent to discarding the continuation generator $\cgen_2$. This results in $$\operatorname{RS}\{G(s)\} \backslash \cgen_2=\mathbb{C} \times\left\{\cgen_1 \mid \cgen_1^2 \sim 1\right\}\cong\mathbb{C} \times \mathbb{Z}_2.$$
	From a topological perspective, this scenario is equivalent to cutting the snail shell in the leftmost panel of Fig.~\ref{fig:2} along Cut-2 (the blue lines). In this case, the structure of the Riemann surface is shown in the middle panel of Fig.~\ref{fig:2}. The Riemann surface $\overline{\mathrm{RS}}\{G(s)\}$ transforms from a snail shell into a flower with countably infinite petals. These petals are connected exclusively through the receptacle (the point at infinity), with each petal comprising a pair of Riemann spheres $\overline{\mathrm{RS}}\{G(s)\}$-$N^{-}$ and $\overline{\mathrm{RS}}\{G(s)\}$-$(N+1)^{+}$.
	Our analysis focuses on the petal structure encompassing the physical sheet $\overline{\operatorname{RS}}\{G(s)\}$-$\mathrm{I}$. This branch can be continuously deformed into a sphere while preserving the topological structure (as shown in the rightmost panel of Fig.~\ref{fig:2}). We will restrict our attention to this branch in the subsequent analysis. In summary, when neglecting the analytic continuation along Cut-2, the Riemann surface $\overline{\mathrm{RS}}\{G(s)\}$ is diffeomorphic to  the two-dimensional sphere $S^2$:
	\begin{equation}
		\overline{\mathrm{RS}}\{G(s)\}\backslash\cgen_2\cong S^2.
		\label{eq:RS-2PGfunc-S2}
	\end{equation}
	
	\section{Riemann surface of partial-wave scattering matrix}
	\label{sec:T-matrix-RS}
	
	According to unitarity, the two-body partial-wave $T$-matrix $\tm(s)$ for two-body coupled-channel scattering\footnote{For a specific definition of the partial-wave $T$-matrix $\tm(s)$, one may refer to Ref.~\cite{Oller:2019rej}.} satisfies the optical theorem as follows:\footnote{The optical theorem states that the partial-wave amplitude of a two-body scattering process is characterized by a series of cuts beginning at the threshold and extending to positive infinity. These cuts are typically referred to as right-hand cuts. In fact, for a rigorous treatment, the contribution of crossing channels must also be considered. These additional contributions introduce left-hand cuts in the amplitude. The discussion in this paper focuses on the right-hand cut portion.}
	\begin{equation}
		\disc\tm(s)=i\sum_{a=1}^{n_c}~\tm(s)\cdot\varrho_a(s)\cdot\tm^*(s)\,\theta(s-s_a),
		\label{eq:optical-theorem}
	\end{equation}
	where the discontinuity of the matrix $\tm(s)$ is defined as $[\disc \tm(s)]_{a b} \equiv \disc\left[\tm_{a b}(s)\right]$. The matrix indices of $\tm(s)$ enumerate all two-body coupled channels (totaling $n_c$ channels), ordered by increasing threshold energies. The matrix $\varrho_a(s) \equiv 2 \iim_a \rho_a(s)$, where $\left(\iim_a\right)_{b c}=\delta_{a b} \delta_{a c}$, $\delta_{a b}$ is the Kronecker delta, and $\rho_a(s)$ and $s_a$ are the two-body phase space factor and the threshold for channel $a$, respectively. If the elements of $\tm(s)$ are real analytic functions, then $$\tm^*\left(s+i 0^{+}\right)=\tm\left(s-i 0^{+}\right)=\tm(s)-\disc \tm(s).$$ 
	Substituting the above result into Eq.~\eqref{eq:optical-theorem} and comparing it with Eq.~\eqref{eq:disc-f-and-f-1}, one obtains:
	\begin{equation}
		\disc\tm^{-1}(s)=-i\sum_{a=1}^{n_c}\varrho_a(s)\,\theta(s-s_a).
		\label{eq:disc-T-inv}
	\end{equation}
	Recalling Eq.~\eqref{eq:disc-G1}, if we introduce the Green's function matrix as follows: $$\gm(s)\equiv\mathrm{diag}\{G_1(s),G_2(s),\cdots,G_{n_c}(s)\},$$
	where $G_a(s)$ is the Green's function of channel $a$, then $\disc\tm^{-1}(s)=\disc\gm(s)$. This relation directly leads to the following result:
	\begin{equation}
		\tm(s)=\left[\hhm(s)+\gm(s)\right]^{-1},
		\label{eq:t-matrix}
	\end{equation}
	where the matrix $\hhm(s)$ satisfies $\disc\hhm(s)=0$.
	
	To determine the Riemann surface structure $\mathrm{RS}\{\tm(s)\}$, we first compute the continuation kernels $\conk_i \tm(s)$, from which we construct the continuation generators $\cgen_i=1-\conk_i$, and the analytic continuation $\cgen_i \tm(s)$ is obtained. Then, by repeatedly carrying out the analytic continuation of $\cgen_i \tm(s)$, one can obtain the structural information of the Riemann surface $\operatorname{RS}\{\tm(s)\}$. 
	Applying the continuation generator $\cgen_i$ to the T-matrix Eq.~\eqref{eq:t-matrix} via the discontinuity rule \eqref{eq:disc-rule-4} yields:
	\begin{equation}
		\cgen_i\tm(s)=\left[\hhm(s)+\cgen_i\gm(s)\right]^{-1}.
		\label{eq:cgen-tm}
	\end{equation}
	This result indicates that the analytic continuation of $\tm(s)$ is completely determined by that of $\gm(s)$, i.e., $\operatorname{RS}\{\tm(s)\} \cong \operatorname{RS}\{\gm(s)\}$. Therefore, the analysis reduces to studying the analytic continuation of $\gm(s)$. Equation \eqref{eq:disc-T-inv} reveals that the Heaviside $\theta$ functions encoding branch cuts in $\disc \gm(s)$ exhibit pairwise overlaps. We split each $\theta$ function into a sum of 
	boxcar functions, defined as $\theta_{[x,y]}(s)=1$ for $s\in [x,y]$ and 0 otherwise:
	\begin{equation}
		\theta(s-s_a)=\sum_{b=a}^{n_c}\theta_{[s_b,s_{b+1}]}(s)\equiv\sum_{b=a}^{n_c}\theta_b(s),\notag
	\end{equation}
	with $s_{n_c+1}$ set to $+\infty$. Substituting the above equation into Eq.~\eqref{eq:disc-T-inv}, exchanging the order of summation, and after rearrangement, one obtains:
	\begin{equation}
		\disc\gm(s)=-i \sum_{a=1}^{n_c}\theta_a(s)\sum_{b=1}^{a}\varrho_b(s).\notag
	\end{equation}
	From the above equation, the continuation kernel of $\gm(s)$ along Cut-1 is $$\conk_{1 a} \gm(s)=-i \sum_{b=1}^a \varrho_b(s)\qquad (a=1, \cdots, n_c).$$
	The continuation generator $\cgen_{1 a}$ is then defined as $\cgen_{1 a} \equiv 1-\conk_{1 a}$, and the analytic continuation of $\gm(s)$ along Cut-1 is
	\begin{equation}
		\cgen_{1a}\gm(s)=\gm(s)+i\sum_{b=1}^{a}\varrho_b(s).
		\label{eq:cgen-gm}
	\end{equation}
	
	Next, let us consider the analytic continuation of $\cgen_{1 a} \gm(s)$. For simplicity, we assume that the maximum pseudo-threshold across all channels is less than the minimum threshold. 
	According to Eqs.~\eqref{eq:disc-rho} and \eqref{eq:cgen-gm}, one can obtain:
		\begin{align}
			\disc\left[\cgen_{1a}\gm(s)\right]=\,&i\sum_{b=1}^{n_c}\,\theta_b(s)\left[\sum_{c=1}^{\min\{a,b\}}\varrho_c(s)-\right.\notag\\
			&\left.\sum_{c>\min\{a,b\}}^{b}\varrho_c(s)\right]+\text{``Cut-2 terms''}-\notag\\
			&\frac{1}{4}\delta(s)\sum_{b=1}^{a}\iim_b\Delta_b.
			\label{eq:disc-cgen-gm-cut-1}
		\end{align}
		where the ``Cut-2 terms'' in the above formula represent the discontinuity along Cut-2, and $\Delta_b=m_{b1}^2-m_{b2}^2$.
	In the main text, we will not consider the analytic continuation of $\cgen_{1 a} \gm(s)$ along Cut-2, which is not of much interest for physical applications. Interested readers may refer to Appendix~\ref{appx:2PGfunc-ac}. From Eq.~\eqref{eq:disc-cgen-gm-cut-1}, the continuation kernel of $\cgen_{1 a} \gm(s)$ along Cut-1 is:
	\begin{equation}
		\conk_{1b}\left[\cgen_{1a}\gm(s)\right]=i\left[\sum_{c=1}^{\min\{a,b\}}\varrho_c(s)-\sum_{c>\min\{a,b\}}^{b}\varrho_c(s)\right].
		\notag
	\end{equation}
	Then, the analytic continuation of $\cgen_{1a}\gm(s)$ along Cut-1 is:
	\begin{equation}
		\cgen_{1b}\left[\cgen_{1a}\gm(s)\right]=\gm(s)+i\sum_{c>\min\{a,b\}}^{\max\{a,b\}}\varrho_c(s).
		\label{eq:cgen-cgen-gm}
	\end{equation}
	From Eq.~\eqref{eq:cgen-cgen-gm}, for any $a$ and $b$, one has the following relations:
	\begin{align}
		&\cgen_{1b}\left[\cgen_{1a}\gm(s)\right]=\cgen_{1a}\left[\cgen_{1b}\gm(s)\right],\notag\\
		&\cgen_{1a}\left[\cgen_{1a}\gm(s)\right]=\gm(s).
		\label{eq:cgen-relation-gm-cut-1}
	\end{align}
	Thus, the structure of the Riemann surface $\mathrm{RS}\{\gm(s)\}$ is as follows:
	\begin{align}
		\mathrm{RS}\{\gm(s)\}=~&\mathbb{C}\times\big\{\cgen_{11},\cdots,\cgen_{1n_c}\big|\cgen_{1a}^2\sim1,\notag\\
		&~~\quad\quad\cgen_{1a}\cgen_{1b}\sim\cgen_{1b}\cgen_{1a}\,(\text{for any }a,b)\big\}\notag\\
		\cong~&\mathbb{C}\times\mathbb{Z}_2^{n_c}.\notag
	\end{align}
	
	\begin{figure*}
		\centering
		\includegraphics[width=0.9\linewidth]{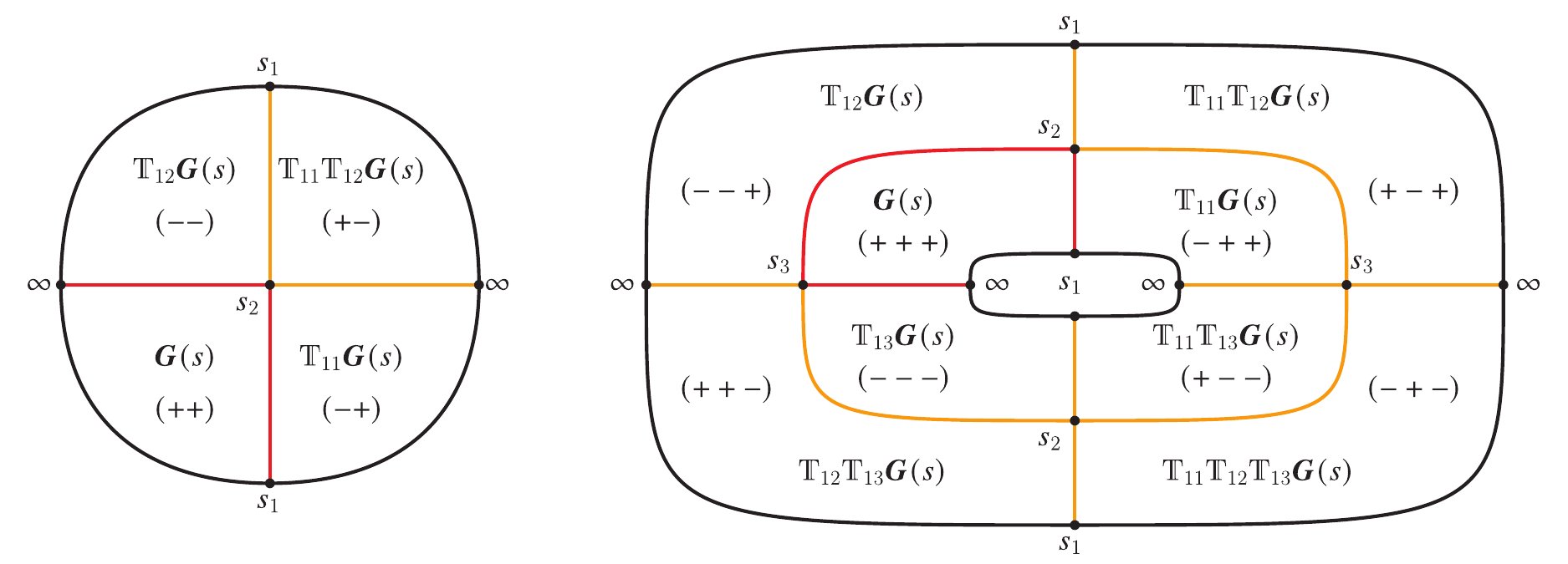}
		\caption{The diagram illustrates the connection patterns among different Riemann sheets following analytic continuation of the partial-wave scattering matrix $\tm(s)$ for $n_c=2$ (left) and $n_c=3$ (right). The red lines indicate Cut-1 on the physical sheet, while the orange lines indicate Cut-1 on unphysical sheets. The black dots indicate the branch points of the Green's functions.}
		\label{fig:3}
	\end{figure*}
	
	The above results indicate that, without considering the analytic continuation along Cut-2, the Riemann surface $\operatorname{RS}\{\gm(s)\}$ consists of $2^{n_c}$ Riemann sheets. The Riemann sheet on which $\gm(s)$ itself lies is designated as the physical sheet. Among all these Riemann sheets, there are $n_c$ Riemann sheets that are directly connected to the physical sheet. 
	The specific forms of the Green's functions on these sheets are given by Eq.~\eqref{eq:cgen-gm}, and these $n_c$ Riemann sheets are referred to as the first-order unphysical sheets, which are closest to the physical sheet. Additionally, there are $C_{n_c}^2$ Riemann sheets connected to the first-order unphysical sheets. 
	The specific forms of the Green's functions on these sheets are given by Eq.~\eqref{eq:cgen-cgen-gm}, and these Riemann sheets are called the second-order unphysical sheets, which are farther from the physical sheet. By analogy, there are $C_{n_c}^m$ $m$-th order unphysical sheets, and the Green's functions on these sheets can be obtained by acting with $m$ different continuation generators $\cgen_a$ on $\gm(s)$. The higher the order $m$, the farther the sheet is from the physical sheet.
	
	To provide a more intuitive understanding of the above conclusion, we present schematic diagrams of the structure of the Riemann surface $\operatorname{RS}\{\gm(s)\}$ for $n_c=2$ and $3$ in Fig.~\ref{fig:3}.\footnote{A similar graphical representation can be found in Ref.~\cite{Yamada:2022xam}, where the eight Riemann sheets in the 3-channel case are organized in a rectangular configuration.}
	In Fig.~\ref{fig:3}, the red lines represent Cut-1 on the physical sheet, while the orange lines represent Cut-1 on the unphysical sheets. Cut-2 has been omitted for simplicity. 
	From Fig.~\ref{fig:3}, one can observe that in the 2-channel case, there are two first-order unphysical sheets ($\cgen_{11}\gm(s)$ and $\cgen_{12}\gm(s)$) directly connected to the physical sheet, as well as one second-order unphysical sheet ($\cgen_{11}\cgen_{12}\gm(s)$). 
	In the 3-channel case, there are three first-order unphysical sheets ($\cgen_{11}\gm(s)$, $\cgen_{12}\gm(s)$, and $\cgen_{13}\gm(s)$) directly connected to the physical sheet, three second-order unphysical sheets ($\cgen_{11}\cgen_{12}\gm(s)$, $\cgen_{12}\cgen_{13}\gm(s)$, and $\cgen_{11}\cgen_{13}\gm(s)$), and one third-order unphysical sheet ($\cgen_{11}\cgen_{12}\cgen_{13}\gm(s)$). 
	Clearly, the higher the order of the unphysical sheet, the farther it is from the physical sheet.
	 
	Moreover, it is evident that all the Riemann sheets in the 3-channel case form a ring-like structure, which is fundamentally different from the topology observed in the 2-channel case. 
	This distinction demonstrates that the Riemann surface $\mathrm{RS}\{\gm(s)\}$ exhibits different genera in the 2-channel and 3-channel scenarios. A detailed discussion of this aspect will follow in Sec.~\ref{sec:uniformization}.

	Poles on different Riemann sheets have distinct effects on the invariant mass distribution within the physical region. 
	Typical resonances correspond to poles on first-order unphysical sheets, with the real part of the pole positioned between the branch points that are the endpoints of the red lines in Fig.~\ref{fig:3}. 
	Such poles are generally closer to the physical sheet and exert a more significant influence on the line shape of the invariant mass distribution within the physical region. 
	If the real part of a pole on the first-order unphysical sheet is not between those branch points represented by the black dots in Fig.~\ref{fig:3}, the pole can only reach the physical region by circling around one threshold, leading to a threshold cusp in the invariant mass distribution. 
	Similar behavior occurs for all poles on higher-order unphysical sheets. 
	For more detailed discussions on how poles on different Riemann sheets affect the invariant mass line shape, readers are referred to, e.g., Refs.~\cite{Dong:2020hxe,Zhang:2024qkg, Nishibuchi:2025uvt}.

	Another commonly used labeling scheme exists. To this end, a new set of continuation generators $\cgen_{1a}'~(a=1,2,\cdots n_c)$ is defined:
	\begin{equation}
		\cgen_{1a}'\equiv\cgen_{1(a-1)}\cgen_{1a},
		\label{eq:def-cgenp-1}
	\end{equation}
	where we stipulate that $\cgen_{10}=1$. According to Eq.~\eqref{eq:cgen-cgen-gm}, the following equations are obtained:
	\begin{align}
		&\cgen_{1a}'\gm(s)=\gm(s)+i\varrho_a(s),\notag\\
		&\cgen_{1a}'\left[\cgen_{1b}'\gm(s)\right]=\cgen_{1b}'\left[\cgen_{1a}'\gm(s)\right],\label{eq:cgenp-gm}\\
		&\cgen_{1a}'\left[\cgen_{1a}'\gm(s)\right]=\gm(s).\notag
	\end{align}
	The physical Riemann sheet is conventionally labeled by the signature ``$++\cdots+$'' ($n_c$ plus signs), where $n_c$ denotes the number of scattering channels. For unphysical sheets corresponding to $\cgen_{1a}'\gm(s)$, we flip the $a$-th positive sign in the physical sheet labeling, e.g., representing the sheet for $\cgen_{11}'\gm(s)$ as ``$-+\cdots+$''. For sheets associated with $\cgen_{1a}'\cgen_{1b}'\gm(s)~(a\neq b)$, two corresponding signs are flipped, for instance, representing the sheet for $\cgen_{11}'\cgen_{12}'\gm(s)$ as ``$--+\cdots+$''. By analogy, for the Riemann sheet corresponding to the Green's function obtained after applying $m$ different continuation generators $\cgen_{1 a}^{\prime}$ to $\gm(s)$, the corresponding $m$ plus signs in the label of the physical sheet are changed to minus signs. 

	In this way, all $2^{n_c}$ Riemann sheets can be labeled, and the expressions of the corresponding Green's functions are given by:
	\begin{equation}
		\gm^{(\bm{k})}(s)\equiv\gm(s)+i\sum_{a=1}^{n_c}k_a\varrho_a(s),
		\label{eq:gm-any-sheet}
	\end{equation}
	where $\bm k=\left(k_1, k_2, \cdots, k_{n_c}\right) \in \mathbb{Z}_2^{n_c}$, and the Green's function on the physical sheet is $\gm(s)=\gm^{(\bm 0)}(s)$.
	
	In summary, we have derived the expressions for $\gm(s)$ on each Riemann sheet after analytic continuation, as well as the structural information of the Riemann surface $\operatorname{RS}\{\gm(s)\}$. Subsequently, based on Eq.~\eqref{eq:cgen-tm}, one can obtain the expressions for the partial-wave scattering matrix $\tm(s)$ on each Riemann sheet after analytic continuation:
	\begin{equation}
		\tm^{(\bm{k})}(s)=\left[\hhm(s)+\bm{G}^{(\bm{k})}(s)\right]^{-1},\notag
	\end{equation}
	together with the structural information of the Riemann surface $\mathrm{RS}\{\tm(s)\}\cong\mathbb{C}\times\mathbb{Z}_2^{n_c}$.
	This equation can also be written equivalently as:
	$$\tm^{(\bm{k})}(s)=\tm(s)\cdot\left[\bm{1}+\varrho^{(\bm{k})}(s)\cdot\tm(s)\right]^{-1},$$
	where $\varrho^{(\bm{k})}(s)=\bm{G}^{(\bm{k})}(s)-\bm{G}(s)$ is defined in Eq.~\eqref{eq:gm-any-sheet}. 

	\begin{figure*}
		\centering
		\includegraphics[width=0.9\linewidth]{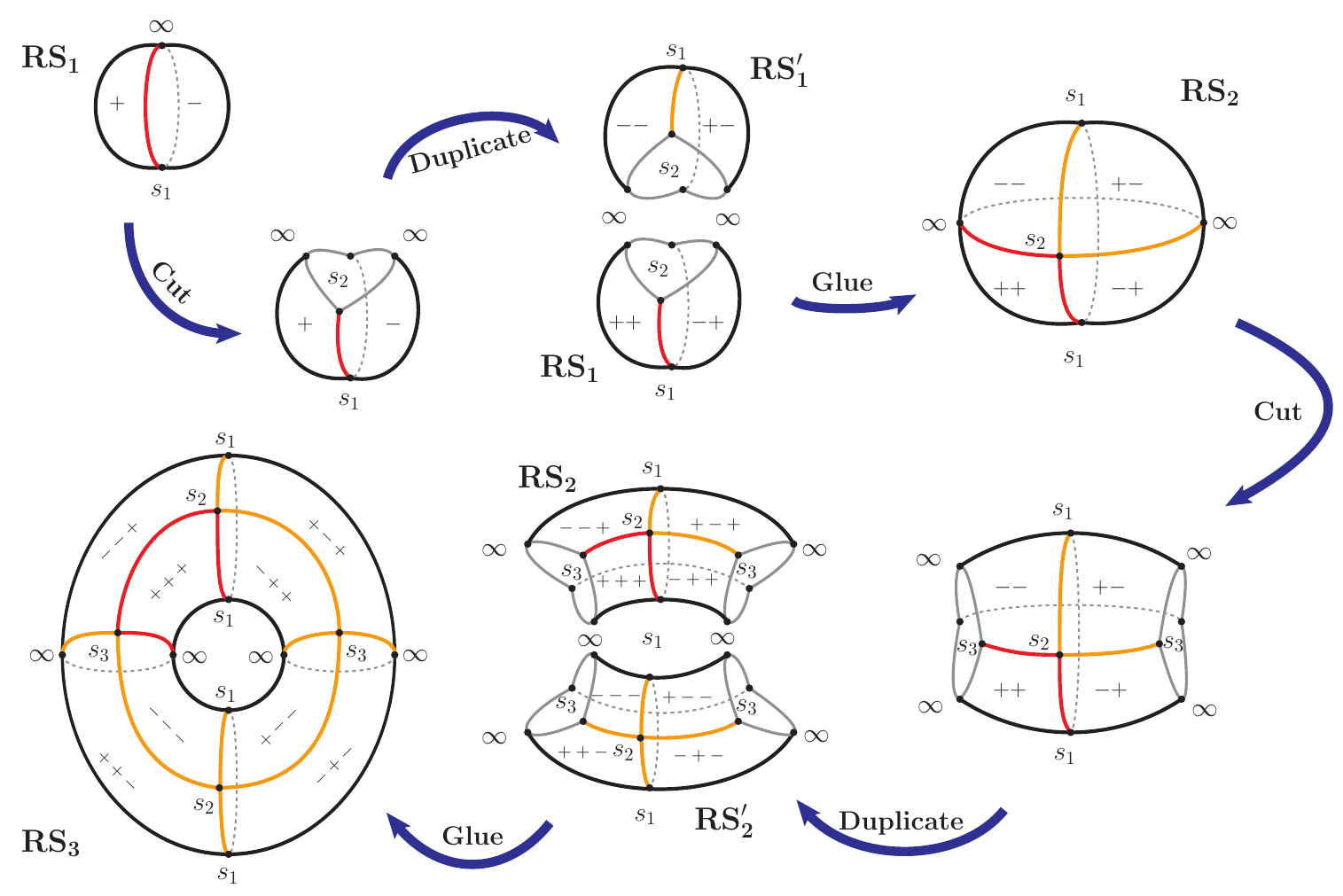}
		\caption{An intuitive schematic diagram illustrating the construction of the Riemann surface $\mathrm{RS}_{1}\to\mathrm{RS}_{2}\to\mathrm{RS}_{3}$ through three steps: cutting, duplicating, and gluing. The red lines indicate Cut-1 on the physical sheet, while the orange lines indicate Cut-1 on the unphysical sheet. The black dots indicate the branch points of the partial-wave scattering matrix $\tm(s)$.}
		\label{fig:4}
	\end{figure*}

	\section{Uniformization of the partial-wave scattering matrix}
	\label{sec:uniformization}

	In this section, let us briefly discuss the uniformization problem of the partial-wave scattering matrix $\tm(s)$---that is, how to map the $2^{n_c}$ Riemann sheets involved in the Riemann surface $\operatorname{RS}\{\tm(s)\}$ to the interior of the Riemann sphere $\hat{\mathbb{C}}$ through a conformal mapping. To this end, one needs to discuss the topological structure of $\overline{\operatorname{RS}}\{\tm(s)\}$ to obtain the genus of the Riemann surface $\overline{\operatorname{RS}} \{\tm(s)\}$. We denote the Riemann surface $\overline{\operatorname{RS}}\{\tm(s)\}$ as $\mathrm{RS}_{n_c}$ for brevity, and its genus as $g_{n_c}$.
	
	First, according to Eq.~\eqref{eq:RS-2PGfunc-S2}, one has $\mathrm{RS}_1\simeq S^2$, and thus $g_1=0$. The structure of $\mathrm{RS}_2$ can be obtained through the following three steps:
	\begin{itemize}
		\item Cutting: On $\mathrm{RS}_1$, start from the threshold $s_2$ on the two Riemann sheets labeled ``$+$'' (physical sheet) and ``$-$'' (unphysical sheet), and make a cut along the real axis.
		\item Duplicating: Add a plus sign ``$+$'' to the label of each Riemann sheet on $\mathrm{RS}_1$. Then, duplicate $\mathrm{RS}_1$ and flip the labels of all Riemann sheets on the duplicate $\mathrm{RS}_1^{\prime}$.
		\item Gluing: Glue $\mathrm{RS}_1$ and $\mathrm{RS}_1^{\prime}$ together according to the connection pattern specified by the analytic continuation to construct the higher-order sheet $\mathrm{RS}_2$.
	\end{itemize}
	An intuitive representation of these steps is shown in Fig.~\ref{fig:4}. As illustrated therein, $\mathrm{RS}_2 \simeq S^2$ and $g_2=0$. By repeating the cutting, duplication, and gluing procedure, one can generate the higher-order Riemann surface $\mathrm{RS}_{n_c+1}$ from the Riemann surface $\mathrm{RS}_{n_c}$. The process of obtaining $\mathrm{RS}_3$ from $\mathrm{RS}_2$ is also depicted in Fig.~\ref{fig:4}. From Fig.~\ref{fig:4}, one can observe that $\mathrm{RS}_3$ is a torus, $\mathrm{RS}_3 \simeq S^1\times S^1$, and thus $g_3=1$. 
	
	Based on this iterative construction process for $\mathrm{RS}_{n_{c}}$, one can derive a recurrence relation for the genera. The genus $g_{n_c}$ can be interpreted as the number of holes on the Riemann surface $\mathrm{RS}_{n_c}$. When constructing $\mathrm{RS}_{n_c+1}$, the total number of holes arises from two sources:
	\begin{enumerate}
		\item[(1)] During duplication, the number of holes on $\mathrm{RS}_{n_c}$ is doubled. This means that each existing hole on $\mathrm{RS}_{n_c}$ gives rise to two corresponding holes on $\mathrm{RS}_{n_c+1}$.
		\item[(2)] During the cutting procedure, $2^{n_c-1}$ cuts are introduced on $\mathrm{RS}_{n_c}$. These cuts contribute $2^{n_c-1}-1$ additional holes during the subsequent gluing process. This contribution accounts for the newly created holes due to the separation and reconnection of the surface. 
	\end{enumerate}
	Combining both contributions, one has
	\begin{equation}
		g_{n_c+1}=2g_{n_c}+2^{n_c-1}-1.\notag
	\end{equation}
	
	From the above recurrence relation and the initial condition $g_1=0$, one can derive the following general formula for the genus:
	\begin{equation}
		g_{n_c}=(n_c-3)2^{n_c-2}+1.
		\label{eq:genus-RS-tm-res}
	\end{equation}
    Our approach concisely reproduces the result in Ref.~\cite{Weidenmuller:1964fbf} for the genus formula of the Riemann surface $\mathrm{RS}_{n_c}$, which will play a pivotal role in the uniformization of the scattering matrix $\tm(s)$.

	Through the above analysis, one can reduce the uniformization problem of $\tm(s)$ to the following mathematical question: 
	Does there exist a conformal mapping that can map a compact Riemann surface of genus $g_{n_c}$ to the Riemann sphere $\hat{\mathbb{C}}$ or its subdomain? If such a mapping exists, what are the characteristics of this conformal mapping? 
	
	This question is answered by the {\bf uniformization theorem} (see, e.g., Ref.~\cite{schlag2014course}) in complex analysis: Every simply connected Riemann surface is conformally equivalent to one of the three canonical Riemann surfaces: the open unit disk $\mathbb{D}$, the complex plane $\mathbb{C}$, or the Riemann sphere $\hat{\mathbb{C}}$.
	This theorem generalizes the Riemann mapping theorem from simply connected open subsets of the complex plane to arbitrary simply connected Riemann surfaces. In particular, for any closed orientable Riemann surface, the corresponding universal cover is determined as follows:
	\begin{itemize}
		\item Genus $g=0$: The Riemann sphere $\hat{\mathbb{C}}$ serves as the universal cover, and the corresponding conformal mapping is given by rational functions.
		\item Genus $g=1$: The complex plane $\mathbb{C}$ serves as the universal cover, and the corresponding conformal mapping is given by elliptic functions.
		\item Genus $g \geq 2$: The unit disk $\mathbb{D}$ serves as the universal cover, and the corresponding conformal mapping is given by automorphic functions.
	\end{itemize}
	
	Therefore, according to Eq.~\eqref{eq:genus-RS-tm-res}, for the two-channel case, we have $g_2=0$, and the four Riemann sheets of $\operatorname{RS}\{\tm(s)\}$ can be mapped onto the Riemann sphere $\hat{\mathbb{C}}$ via rational functions. 
	For the specific construction of such a mapping, see Ref.~\cite{Kato:1965iee}. 
	For the three-channel case, we have $g_3=1$, and the eight Riemann sheets of $\operatorname{RS}\{\tm(s)\}$ can be mapped onto the complex plane $\mathbb{C}$ via elliptic functions with two periods. 
	The specific construction of such a mapping was recently carried out in Ref.~\cite{Yamada:2022xam}. For cases where the number of channels satisfies $n_c>3$, we have $g_{n_c}\geq 5$, and the $2^{n_c}$ Riemann sheets of $\operatorname{RS}\{\tm(s)\}$ can be mapped onto the unit disk $\mathbb{D}$ via automorphic functions.

	\section{Summary}
	\label{sec:summary}
	
	In this work, we have proposed a novel method, called the discontinuity calculus, for calculating the analytic continuation and Riemann surface structure of complex functions. This method has been applied to coupled-channel problems in two-body scattering, providing a systematic analysis of the analytic continuation and Riemann surface structure of two-body coupled-channel scattering matrices. 
	Furthermore, we have rederived the genus formula for the Riemann surfaces associated with two-body coupled-channel scattering matrices and established a connection with the uniformization theorem in complex analysis. Our findings are consistent with the established uniformization mapping constructions for two- and three-channel systems, as demonstrated in Refs.~\cite{Kato:1965iee,Yamada:2022xam}.
	
	The discontinuity calculus introduced here can also be extended to calculate discontinuities and perform analytic continuation of multivariable complex functions. This extension is applicable for analyzing the analytic continuation and Riemann surface structure of scattering matrices in multi-body (three-body and beyond) coupled-channel scattering problems. Such an extension may provide new insights for analyzing coupled-channel problems involving three-body effects.

	\bigskip
	\begin{acknowledgments}
		We thank useful discussions with Bing Wu.
		This work is supported in part by the National Natural Science Foundation of China (NSFC) under Grants No.~12405100, No.~12125507, No.~12361141819, and No.~12447101; by the Chinese Academy of Sciences under Grant No.~YSBR-101; by the Postdoctoral Fellowship Program of China Postdoctoral Science Foundation under Grant No.~GZC20232773 and No.~2023M74360; and by the National Key R\&D Program of China under Grant No. 2023YFA1606703. 
	\end{acknowledgments}
	
	\appendix
	
	\section{Alternative definition of the discontinuity calculus}
	\label{appx:def-dgen}

	Define the operator $\dgen\equiv1-\disc$. The four properties (\ref{eq:disc-rule-1}--\ref{eq:disc-rule-4}) satisfied by the discontinuity operator $\disc$ introduced in Sec.~\ref{sec:def-disc} can be equivalently expressed as the following four properties satisfied by the operator $\dgen$:

	\noindent (1) Preservation of holomorphy:
	\begin{equation}
		\dgen h(z)=h(z).\tag{R1$'$}
	\end{equation}
	(2) Linearity:
	\begin{equation}
		\dgen\left[\alpha_1 f_1(z)+\alpha_2 f_2(z)\right]=\alpha_1\dgen f_1(z)+\alpha_2\dgen f_2(z).\tag{R2$'$}
	\end{equation}
	(3) Associative homomorphism:  
	\begin{equation}
		\dgen \left[f_1(z)f_2(z)\right]=\dgen f_1(z) \dgen f_2(z).\tag{R3$'$}
	\end{equation}
	(4) Chain rule:
	\begin{equation}
		\dgen F\left[f(z)\right]=\dgen F\left[\omega\right]\big|_{\omega=\dgen f(z)}.\tag{R4$'$}
	\end{equation}

	\section{Direct calculation of discontinuities in the two-point Green's function}
	\label{appx:disc-2PGfunc-def}
	
	The two-point Green's function $G(s)$ is defined by Eq.~\eqref{eq:def-2PGreenfunction}, which can be rewritten as:
	\begin{equation}
		G(s)=\int_{\mathbb{R}^4} \frac{i d^4 q}{(2 \pi)^4} \frac{1}{D_1^+ D_2^- D_3^- D_4^+},
        \label{eq:def-G-mod}
	\end{equation}
	where 
	\begin{align}
		& D_1^\pm\equiv q^0+E / 2-\omega_1\pm i \epsilon,\notag\\
		& D_2^\pm\equiv q^0+E / 2+\omega_1\pm i \epsilon, \notag\\
		& D_3^\pm\equiv q^0-E / 2+\omega_2\pm i \epsilon, \notag\\
		& D_4^\pm\equiv q^0-E / 2-\omega_2\pm i \epsilon. \notag
	\end{align}
	The calculation is performed in the center-of-mass frame by setting $p=(E,\mathbf{0})$, where $\omega_i=\sqrt{|\mathbf{q}|^2+m_i^2}~(i=1,2)$ denotes the on-shell energy of particle-$i$, and $E=\sqrt{s}$ represents the total energy.
	
	We now consider the discontinuity of $G(s)$ with respect to $E$:
	\begin{equation}
		\disc_{E} G(s)=\int_{\mathbb{R}^4} \frac{i d^4 q}{(2 \pi)^4} \disc_{E}\left(\frac{1}{D_1^+ D_2^- D_3^- D_4^+}\right).
		\label{eq:disc-G-def}
	\end{equation}
	According to Eq.~\eqref{eq:disc-rule-3}, we need to compute the discontinuities of all four terms in the integrand. By convention, the physical region corresponds to the upper edge of the real axis in the complex $E$-plane. Using Eq.~\eqref{eq:disc-1/x}, we obtain:
	\begin{align}
		& \disc_{E}\left(\frac{1}{D_1^\pm}\right)=-2 \pi i ~\delta\!\left(q^0+E / 2-\omega_1\right) \equiv-2 \pi i~ \delta_1, \notag\\
		& \disc_{E}\left(\frac{1}{D_2^\pm}\right)=-2 \pi i ~\delta\!\left(q^0+E / 2+\omega_1\right) \equiv-2 \pi i~ \delta_2, \notag\\
		& \disc_{E}\left(\frac{1}{D_3^\pm}\right)=+2 \pi i ~\delta\!\left(q^0-E / 2+\omega_2\right) \equiv +2 \pi i~ \delta_3, \notag\\
		& \disc_{E}\left(\frac{1}{D_4^\pm}\right)=+2 \pi i ~\delta\!\left(q^0-E / 2-\omega_2\right) \equiv +2 \pi i~ \delta_4.\notag
	\end{align}
	Since the small imaginary component $\pm i\epsilon$ in the argument of the Dirac delta function does not affect subsequent analyses, we have omitted it in the results above for simplicity. Substituting the above results into Eq.~\eqref{eq:disc-G-def} yields:
\begin{widetext}
		\begin{align}
		\disc_{E}G(s)=\int_{\mathbb{R}^4} \frac{i d^4 q}{(2 \pi)^4} & \left[~~2 \pi i\left(-\frac{\delta_1}{D_2^- D_3^- D_4^+}-\frac{\delta_2}{D_1^+ D_3^- D_4^+}+\frac{\delta_3}{D_1^+ D_2^- D_4^+}+\frac{\delta_4}{D_1^+ D_2^- D_3^-}\right)-\right. \notag\\
		& \quad\, (2 \pi i)^2\left(\frac{\delta_1 \delta_2}{D_3^- D_4^+}-\frac{\delta_1 \delta_3}{D_2^- D_4^+}-\frac{\delta_1 \delta_4}{D_2^- D_3^-}-\frac{\delta_2 \delta_3}{D_1^+ D_4^+}-\frac{\delta_2 \delta_4}{D_1^+ D_3^-}+\frac{\delta_3 \delta_4}{D_1^+ D_2^-}\right)+ \notag\\
		& \quad\left.(2 \pi i)^3\left(\frac{\delta_1 \delta_2 \delta_3}{D_4^+}+\frac{\delta_1 \delta_2 \delta_4}{D_3^-}-\frac{\delta_1 \delta_3 \delta_4}{D_2^-}-\frac{\delta_2 \delta_3 \delta_4}{D_1^+}\right)-(2 \pi i)^4 \delta_1 \delta_2 \delta_3 \delta_4~~\right].
		\label{eq:disc-G-expand}
	\end{align}
	%
	
	In this work, we restrict our analysis to scenarios with non-zero masses $m_1,m_2>0$, ensuring that the on-shell energies $\omega_1,\omega_2>0$. This mass condition leads to the vanishing products $\delta_1\delta_2=\delta_3\delta_4=0$. Substituting this condition into Eq.~\eqref{eq:disc-G-expand}, we obtain:
		\begin{align}
			\disc_{E}G(s)=\int_{\mathbb{R}^4} \frac{id^4 q}{(2 \pi)^4}&\left[~~2 \pi i\left(-\frac{\delta_1}{D_2^- D_3^- D_4^+}-\frac{\delta_2}{D_1^+ D_3^- D_4^+}+\frac{\delta_3}{D_1^+ D_2^- D_4^+}+\frac{\delta_4}{D_1^+ D_2^- D_3^-}\right)+\right.\notag\\
			&\quad\left.(2 \pi i)^2\left(\frac{\delta_1 \delta_3}{D_2^- D_4^+}+\frac{\delta_1 \delta_4}{D_2^- D_3^-}+\frac{\delta_2 \delta_3}{D_1^+ D_4^+}+\frac{\delta_2 \delta_4}{D_1^+ D_3^-}\right)~~\right].\notag
		\end{align}
	\end{widetext}
	The first line on the right-hand side contains terms with a single Dirac-$\delta$ function, referred to as single-pole terms, while the second line comprises terms with two Dirac-$\delta$ functions, referred to as double-pole terms.

    \begin{figure*}
		\centering
		\includegraphics[width=1\linewidth]{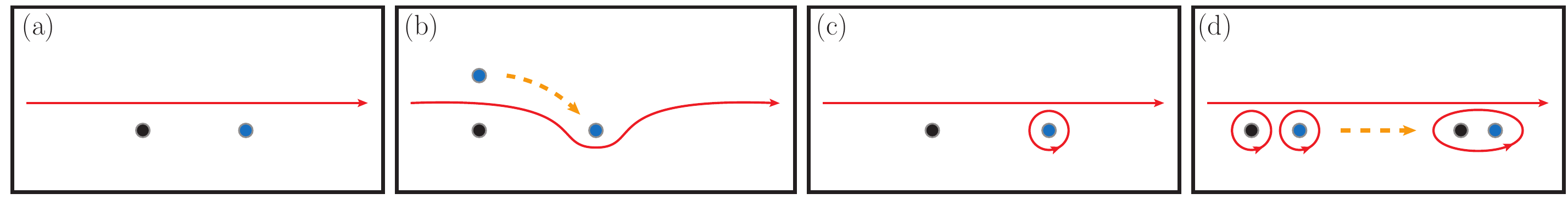}
		\caption{Illustrative explanation of the last equality in Eq.~\eqref{eq:disc-I}. The red lines represent integration contours, with arrowheads indicating the integration direction. The black and blue dots denote the two poles in the integrand.
		(a) The contour avoids being pinched when both singularities coalesce and lie in the same half-plane. As a result, no singularities emerge in the integral.
		(b) If the poles are positioned on opposite sides of the contour, their coalescence pinches the contour, leading to singularities in the integral.
		(c) By deforming the contour in panel (b), the two poles can be relocated to the same side (e.g., by moving the blue pole from the upper to the lower half-plane). To preserve the continuity of the integral, an additional integration over a closed contour with winding number $+1$ encircling the blue pole is introduced. Even in this configuration, pole coalescence continues to pinch the contour, thereby generating singularities.
		(d) When the poles approach each other while being enclosed by sub-contours with identical winding numbers (e.g., both with winding number $+1$), no pinching occurs, and the integral remains free of singularities.}
		\label{fig:5}
	\end{figure*}
    
	We first analyze the single-pole terms. For these terms, branch points that contribute to the discontinuity can emerge in the integral only when the pole positions coincide with the endpoints of the complex $q^0$-plane integration contour. In our case, however, the integration path spans the entire real axis, which can be equivalently viewed as a closed contour that closes at infinity, and thus inherently lacks endpoints. Consequently, the single-pole terms do not contribute to the discontinuity.
	
	Next, we investigate the contributions from double-pole terms. For clarity, we henceforth denote the relevant poles directly as $\delta_i(i=1,2,3,4)$. Consider first a generalized integral of the following form:
	\begin{equation}
		I(s,\bm{n})\equiv\int_{\mathcal{C}(\bm{n})\times\mathbb{R}^3}\frac{i d^4 q}{(2 \pi)^4}\frac{1}{D_1^+ D_2^+ D_3^+ D_4^+},\notag
	\end{equation}
	where $\mathcal{C}(\bm{n})$ represents the integration contour in the complex $q^0$-plane, with the vector $\bm{n}=(n_1,n_2,n_3,n_4)\in\mathbb{Z}^4$ labeling the homotopy equivalence classes of closed contours. Specifically, for $\bm{n}=\mathbf{0}$, the contour $\mathcal{C}(\mathbf{0})$ coincides with the entire real axis. For $\bm{n}\neq\mathbf{0}$, the contour $\mathcal{C}(\bm{n})$ comprises not only the real axis but also closed loops encircling each pole $\delta_i$ with winding number $n_i$, where $n_i>0$ indicates counterclockwise encirclements and $n_i<0$ indicates clockwise ones.
	
	From the analysis above, we conclude that only the double-pole terms contribute to the discontinuity of the integral $I(s,\bm{n})$. For a specific double-pole term, its contribution depends on whether the integration contour is pinched between the two corresponding poles. In general, we have:
	\begin{widetext}
		\begin{align}
			\disc_E I(s,\bm{n})=&\int_{\mathcal{C}(\bm{n})\times\mathbb{R}^3}\frac{i d^4 q}{(2 \pi)^4}\left[(2 \pi i)^2\left(\frac{\delta_1 \delta_3}{D_2^+ D_4^+}+\frac{\delta_1 \delta_4}{D_2^+ D_3^+}+\frac{\delta_2 \delta_3}{D_1^+ D_4^+}+\frac{\delta_2 \delta_4}{D_1^+ D_3^+}\right)\right] \notag\\
            =&\int_{\mathbb{R}^4}\frac{i d^4 q}{(2 \pi)^4}\left\{\left(2\pi i\right)^2\left[c_{13}(\bm{n})\frac{\delta_1\delta_3}{D_2^+D_4^+}+c_{14}(\bm{n})\frac{\delta_1\delta_4}{D_2^+D_3^+}+c_{23}(\bm{n})\frac{\delta_2\delta_3}{D_1^+D_4^+}+c_{24}(\bm{n})\frac{\delta_2\delta_4}{D_1^+D_3^+}\right]\right\},
			\label{eq:disc-I}
		\end{align}
	where $c_{ij}(\bm{n})~(ij\in\{13,14,23,24\})$ are integer-valued functions of the homotopy class $\bm{n}$, explicitly given by $c_{ij}(\bm{n})=n_i-n_j$. 
    For an intuitive explanation of this result, we refer to Fig.~\ref{fig:5}.

	Crucially, the two-point Green's function $G(s)$ in Eq.~\eqref{eq:def-G-mod} specifies that two of the four poles ($\delta_2$ and $\delta_3$) are located on the upper edge of the complex $q^0$-plane integration contour. By deforming the contour to relocate all poles to the lower edge of the integration contour, we acquire an additional integration over closed loops encircling $\delta_2$ and $\delta_3$ with winding number $+1$. Thus, by setting $\bm{n}_G=(0,1,1,0)$, we identify $G(s)=I(s,\bm{n}_G)$. Substitution into Eq.~\eqref{eq:disc-I} yields:
		\begin{align}
			\disc_E G(s) =\disc_E I(s,\bm{n}_G)=& \int_{\mathbb{R}^4}\frac{i d^4 q}{(2 \pi)^4}\left[\left(2\pi i\right)^2\left(-\frac{\delta_1\delta_3}{D_2^+D_4^+}+\frac{\delta_2\delta_4}{D_1^+D_3^+}\right)\right]\notag\\
			= & \int_{\mathbb{R}^4} \frac{i d^4 q}{(2 \pi)^4}\frac{( 2 \pi i ) ^ { 2 }}{4\omega_1\omega_2} \left[\delta\!\left(E -\omega_1-\omega_2\right) \delta\!\left(q^0-E/2+\omega_2\right)-\delta\!\left(E+\omega_1+\omega_2\right) \delta\!\left(q^0-E/2-\omega_2\right)\right] \notag\\
			= & -2 i \rho(E^2)\left[\theta\left(E-m_1-m_2\right)-\theta\left(-E-m_1-m_2\right)\right] \notag\\
			= & -2 i \rho(s)\,\theta\left(s-s_{+}\right).\notag
		\end{align}
	This result is completely consistent with the discontinuity formula derived in Eq.~\eqref{eq:disc-G1}.
    \end{widetext}
	
	\section{Complete analytic continuation of the Green's function matrix}
	\label{appx:2PGfunc-ac}
	
	The ``Cut-2 terms'' in Eq.~\eqref{eq:disc-cgen-gm-cut-1} are given by:
	\begin{align}
		\text{``Cut-2 terms''}=i\sum_{b=1}^{n_c}\theta_b'(s)\sum_{1\leq c\leq a}^{p(c)\leq b}\varrho_c(s),\notag
	\end{align}
	where $\theta_b^{\prime}(s) \equiv \theta_{[s_b', s_{b+1}']}(s)$,
	$s_a^{\prime}$ $(a=1, \cdots, n_c)$ are the pseudo-thresholds arranged in descending order, $p(c)$ denotes the order of the pseudo-threshold of channel $c$, and we set $s_{n_c+1}^{\prime} $ to $-\infty$. 
	
	We have discussed the analytic continuation of $\cgen_{1 a} \gm(s)$ along Cut-1 in the main text, as shown in Eq.~\eqref{eq:cgen-cgen-gm}. Besides Cut-1, $\cgen_{1 a} \gm(s)$ has another cut, Cut-2, which does not exist on the physical sheet but exists on all other sheets. Thus, one needs to introduce new continuation kernels $\conk_{2 a}$ $(a=1, \cdots, n_c)$, which satisfy the following equations:
	\begin{align}
		&\conk_{2a}\gm(s)=0,\notag\\
		&\conk_{2b}\left[\cgen_{1a}\gm(s)\right]=i\sum_{1\leq c\leq a}^{p(c)\leq b}\varrho_c(s),\notag
	\end{align}
	and the continuation generators $\cgen_{2a}=1-\conk_{2a}$:
	\begin{align}
		&\cgen_{2a}\gm(s)=\gm(s),\label{eq:cgen-gm-cut-2}\\
		&\cgen_{2b}\left[\cgen_{1a}\gm(s)\right]=\gm(s)-i\sum_{1\leq c\leq a}^{p(c)\leq b}\varrho_c(s)+i\sum_{1\leq c\leq a}^{p(c)> b}\varrho_c(s).\notag
	\end{align}
	According to Eq.~\eqref{eq:cgen-gm}, one can obtain:
	\begin{align}
		\cgen_{1a}\left[\cgen_{2b}\gm(s)\right]=~&\cgen_{1a}\gm(s)=\gm(s)+i\sum_{b=1}^a\varrho_b(s)\notag\\
		\neq~&\cgen_{2b}\left[\cgen_{1a}\gm(s)\right].\notag
	\end{align}
	This result indicates that the actions of the continuation generators $\cgen_{1a}$ and $\cgen_{2b}$ on $\gm(s)$ do not commute. 
	Furthermore, the analytic continuation of $\cgen_{2b}\left[\cgen_{1a}\gm(s)\right]$ across Cut-2 can be evaluated to yield:
	\begin{align}
		\cgen_{2c}\left\{\cgen_{2b}\left[\cgen_{1a}\gm(s)\right]\right\}=~&\gm(s)+i\sum_{1\leq d\leq a}^{p(d)\leq\min\{b,c\}}\varrho_d(s)~-\notag\\
		&i\sum_{1\leq d\leq a}^{\min\{b,c\}<p(d)\leq\max\{b,c\}}\varrho_d(s)~+\notag\\
		&i\sum_{1\leq d\leq a}^{p(d)>\max\{b,c\}}\varrho_d(s).\notag
	\end{align}
	From this equation, we can derive the following relations:
	\begin{align}
		&\cgen_{2c}\left\{\cgen_{2b}\left[\cgen_{1a}\gm(s)\right]\right\}= \cgen_{2b}\left\{\cgen_{2c}\left[\cgen_{1a}\gm(s)\right]\right\},\notag\\
		&\cgen_{2b}\left\{\cgen_{2b}\left[\cgen_{1a}\gm(s)\right]\right\}= \cgen_{1a}\gm(s),\notag
	\end{align}
	which imply the equivalence relations $\cgen_{2a}\cgen_{2b}\sim\cgen_{2b}\cgen_{2a}$ and $\cgen_{2a}^2\sim1$.
	
	To clarify the structural information of the Riemann surface $\mathrm{RS}\{\gm(s)\}$, we define a new set of continuation generators $\cgen_{2a}'~(a=1,\cdots,n_c)$:
	\begin{equation}
		\cgen_{2a}'\equiv\cgen_{2[p(a)-1]}\cgen_{2p(a)},\notag
	\end{equation}
	where we stipulate that $\cgen_{20}=1$. According to Eq.~\eqref{eq:disc-rho}, one can obtain
	\begin{align}
		&\cgen_{2a}'\varrho_a(s)=-\varrho_a(s),\notag\\
		&\cgen_{2a}'\varrho_b(s)=\varrho_b(s)~(a\neq b).
		\label{eq:cgenp-varrho-1-and-2}
	\end{align}
	If we replace $\cgen_{2 a}^{\prime}$ with $\cgen_{1 a}^{\prime}$ from Eq.~\eqref{eq:def-cgenp-1} in the above equations, they still hold. 
	
	Furthermore, one can define the raising operator $\ppm_a$ and the lowering operator $\ppm_{-a}~(a=1,\cdots,n_c)$:
	\begin{align}
		\ppm_a\equiv\cgen_{1a}'\cgen_{2a}', \quad \ppm_{-a}\equiv\cgen_{2a}'\cgen_{1a}'.\notag
	\end{align}
	Combining Eqs.~(\ref{eq:cgenp-gm},\ref{eq:cgen-gm-cut-2},\ref{eq:cgenp-varrho-1-and-2}), one can obtain
	\begin{align}
		&\ppm_{\pm a}^{k}\gm(s)=\gm(s)\pm k \varrho_a(s),\notag\\
		&\ppm_{\pm a}^{k}\ppm_{\pm b}^{k'}\gm(s)\hspace{-.5mm}=\hspace{-.5mm}\ppm_{\pm b}^{k'}\ppm_{\pm a}^{k}\gm(s)\hspace{-.5mm}=\hspace{-.5mm}\gm(s)\pm k \varrho_a(s)\pm k'\varrho_b(s),\notag
	\end{align}
	where $k,k'\in\mathbb{Z}$. These equations imply that the structure of the Riemann surface $\mathrm{RS}\{\gm(s)\}$ is:
	\begin{align}
		\mathrm{RS}\{\gm(s)\}=~&\mathbb{C}\times\big\{\ppm_{\pm1},\cdots,\ppm_{\pm n_c}\big|\ppm_a\ppm_{-a}\sim1,\notag\\
		&~~\quad\quad\ppm_a\ppm_b\sim\ppm_{b}\ppm_{a}(\text{for any }a,b)\big\}\notag\\
		\cong~&\mathbb{C}\times\mathbb{Z}^{n_c}.\notag
	\end{align} 
	
	In summary, $\mathrm{RS}\{\gm(s)\}$ consists of countably infinitely many Riemann sheets. The corresponding Green's function on each Riemann sheet is given by:
	\begin{equation}
		\gm^{(\bm{k})}(s)\equiv\gm(s)+i\sum_{a=1}^{n_c}k_a\varrho_a(s),\notag
	\end{equation}
	where $\bm{k}=(k_1,k_2,\cdots,k_{n_c})\in\mathbb{Z}^{n_c}$.

	\bibliography{refs.bib}

\end{document}